# A 3.0MeV KOMAC/KTF RFQ Linac

J.M. Han, K.S. Kim, Y.J. Kim, M.Y. Park, Y.S. Cho, B.H. Choi, KAERI, Taejon, Korea
Y.S. Bae, K.Y. Shim, K.H. Kim, I.S. Ko, POSTECH, Pohang, Korea
S.J. Cheon, KAIST, Taejon, Korea
Y. Oguri, TIT, Japan

*Abstract*

The Radio-Frequency Quadrupole (RFQ) linac that will accelerate a 20mA proton beam from 50keV to 3MeV has been designed and is being fabricated as the first phase, KOMAC Test Facility (KTF), of the Korea Multipurpose Accelerator Complex (KOMAC) project at the Korea Atomic Energy Research Institute (KAERI). The physical, engineering designs and fabrication status of the RFQ are described.

## 1 INTRODUCTION

The linear accelerator for the KOMAC Project [1] will include a 3MeV, 350MHz cw RFQ linac. The KOMAC/KTF RFQ [2,3] concept is shown in Fig. 1 with the main parameters given in table 1. The KTF RFQ bunches, focuses, and accelerates the 50keV $H^+/H^-$ beams and derives a 3.0MeV beam at its exit. The RFQ is a 324cm-long, 4-vanes type and consists of 56 tuners, 16 vacuum ports, 1 coupling plate, 4 rf drive couplers, 96 cooling passages, and 8 stabiliser rods. The RFQ is machined into OFH-Copper, will be integrated from four separate 81cm-long sections which are constructed by using vacuum furnace brazing. RF power is supplied to the RFQ which operates at 100% duty-factor by one klystron of 1MW.

The physics and engineering design study of the KTF RFQ linac are presented in section 2. Section 3 describes the 450keV RFQ which is a test stand for the development of the KTF RFQ linac. Section 4 presents the present status of the KTF RFQ.

Table 1. The KOMAC/KTF RFQ Linac Parameters.

| PARAMETER | VALUE |
|---|---|
| Operating frequency | 350 MHz |
| Particles | $H^+ / H^-$ |
| Input / Output Current | 21 / 20 mA |
| Input / Output Energy | 0.05 / 3.0 MeV |
| Input / Output Emittance, Transverse/norm. | 0.02 /0.023 π-cm-mrad rms |
| Output Emittance, Longitudinal | 0.246 MeV-deg |
| Transmission | 95 % |
| RFQ Structure Type | 4-vanes |
| Duty Factor | 100 % |
| Peak Surface Field | 1.8 Kilpatrick |
| Structure Power | 350 kW |
| Beam Power | 68 kW |
| Total Power | 418 kw |
| Length | 324 cm |

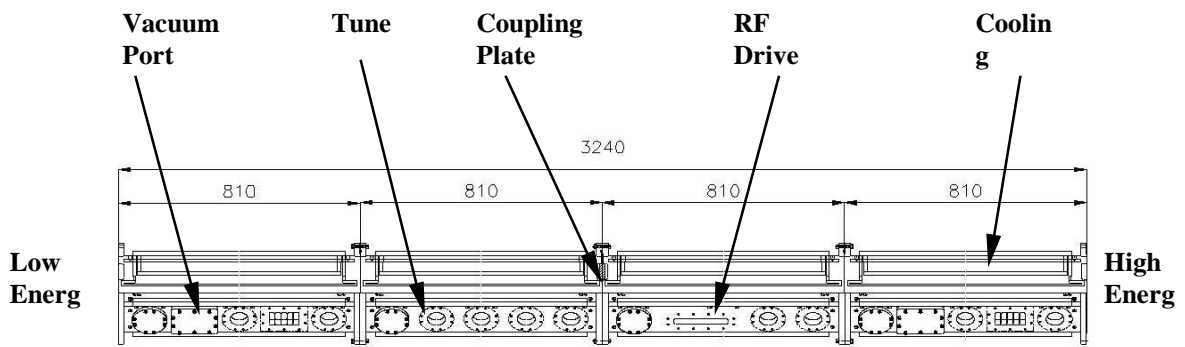

Figure 1. 3MeV, 350MHz, cw KOMAC/KTF RFQ

## 2  3.0MeV RFQ LINAC

### 2.1  Cavity Design

The design of the 3MeV RFQ has been completed. In the KTF RFQ design, a main issue is to accelerate the mixing $H^+/H^-$ beam at the same time. The motion of the mixing $H^+/H^-$ beam into the RFQ has been studied by using a time marching beam dynamics code, QLASSI[4]. Fig. 2 shows that the longitudinal beam loss increases with the concentration of negative ions by the bunching process which is distributed by attractive forces when the mixing ratio of $H^-$ is more than 30%. The transverse beam loss decreases with the mixing ratio of $H^-$ by the space charge compensation in the low energy sections.

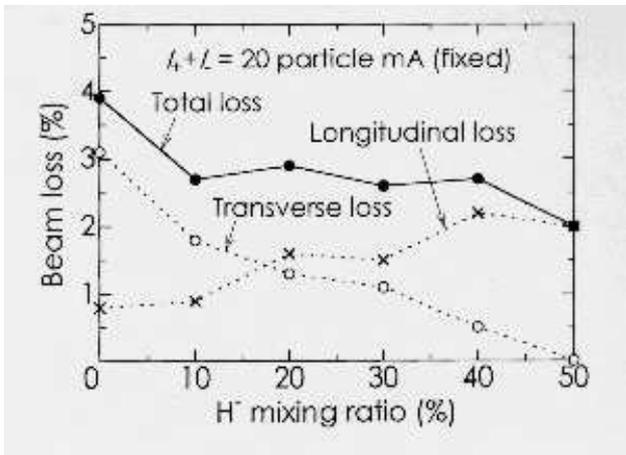

Figure 2. Dependency of the Beam Transmission Rate and $H^-$ Mixing Ratio.

The average RFQ cavity structure power by rf thermal loads is 0.35 MW and the peak surface heat flux on the cavity wall is 0.13 MW/m$^2$ at the high energy end. In order to remove this heat, we consider 24 longitudinal coolant passages in each of the sections. In the design of the coolant passages, we considered the thermal behaviour of the vane during CW operation, the efficiency of cooling and fabricating cost. The thermal and structure analysis was studied with the ANSYS code. Because of the flow erosion of the coolant passages, we considered the maximum allowable velocity of the normal coolant as 4m/sec. From the thermal-structural analysis of ANSYS, the peak temperature on the cavity wall is 51.4 $^\circ$C, the maximum displacement is 42μm and the intensity stress is 13MPa. The temperature of the coolants on the cavity wall varies to maintain the cavity on the resonance frequency.

### 2.2  RF Power System

The total power simulated is 418kW, including beam loading and power dissipation by a cavity wall, when an additional 50% of the power is allowed as the difference between the theoretical model of the RFQ and the real device built. This power is delivered by a single klystron, capable of 1MW. The 350MHz klystron and RF windows will be supplied by Thomson Co. Ltd. The power is coupled in the cavity with a set of four coupling loops. Each port will therefore carry an average rf power of 120kW.

### 2.3  Cold Model

In order to test the fabrication accuracy and validate the simulation studies which were performed by PARMTEQM[5], VANES, SUPERFISH, MAFIA, ANSYS codes, a full size 324cm-long RFQ cold model was made of aluminium as shown in Fig. 3. Though this model does not operate with rf power and under vacuum pumping, it has rf power coupling ports, tuner ports, stabilizer rod, coupling plate, end plates, and vacuum ports which are given by the 3-dim drawing. By tuning the undercut depth and end plates, we obtained the optimum vane-end geometry and the required 350 MHz resonant frequency.

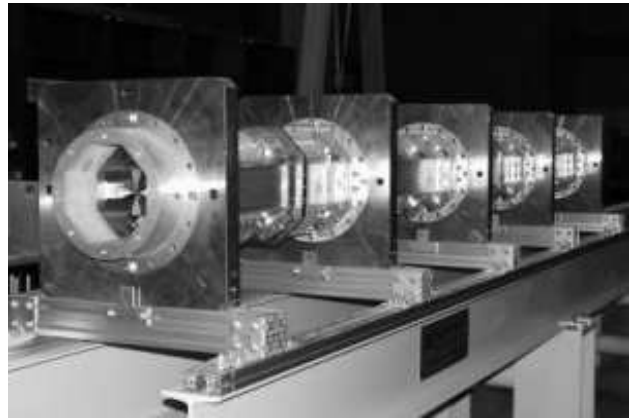

Figure 3. The 3MeV RFQ Cold Model.

## 3  0.45MeV TEST RFQ

A 450keV test RFQ has been designed and fabricated to understand the construction process, cooling, control, rf drive system, and beam diagnostic techniques. Design of the RFQ was done by KAERI and POSTECH, and fabrication was done at Dae-Ung Engineering Company and VITZRO TECH Co., Ltd.

Fig. 4 shows a 96.4cm long 0.45MeV RFQ which was brazed in a vacuum furnace. The RFQ was brazed in a vertical orientation with LUCAS BVag-8, AgCu alloy with a liquid temperature of 780 $^\circ$C. The four quadrants of the RFQ have been fabricated separately and brazed. Thus the RFQ is the completed monolithic structure and the vanes are permanently aligned. This structure serves to mitigate the cost and to simplify the mechanical support system. Because of the leak of a brazing surface and the strain of the RFQ structure by the furnace heat, it is important to determine the appropriate shape of the brazing area. To determine the appropriate shape, we have performed two brazing tests. Testing of the brazed RFQ showed it to be leak-tight. The coolant passages in the cavity wall and vane area was the deep-hole drilled and was brazed in a vacuum furnace.

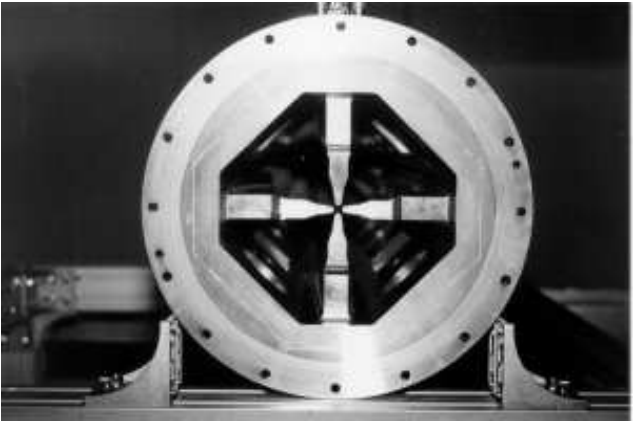

Figure 4. A brazed 0.45MeV RFQ.

The frequency and unloaded Q was measured with a network analyzer in atmosphere, i.e. not under a vacuum. Without tuning, the measured frequency and Q were 349.63MHz and 5300 by observing the 1dB bandwidth, respectively.

Fig. 5 shows the rf tuner which was fabricated to tune the cavity to the operating frequency. By moving four tuners to a 2.5cm inserted position, a total tuning range of 2.5MHz was measured.

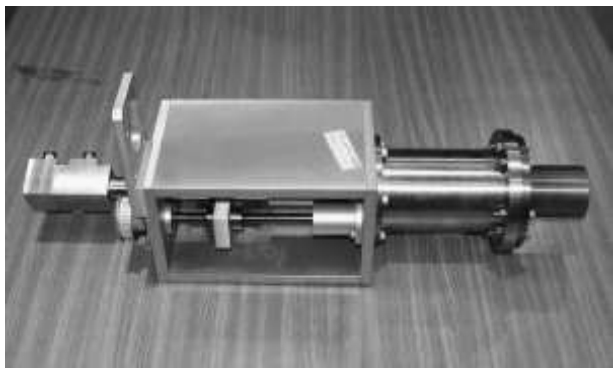

Figure 5. RF tuner.

The KTF RFQ has a rectangular undercut of the vanes. The exact dimension of the undercut has been determined empirically by cutting a vane which was fabricated into the OFHC. Fig. 6 shows the variation of the resonant frequency versus the depth of the undercut. The resonant frequency of the RFQ cavity inversely decreases with undercut depth. To maximise the effect of the stabilizer rod, we determined that the undercut depth and vane to end-plate length are 28mm and 35mm, respectively. In this case, the quadrupole-dipole separation was 10MHz.

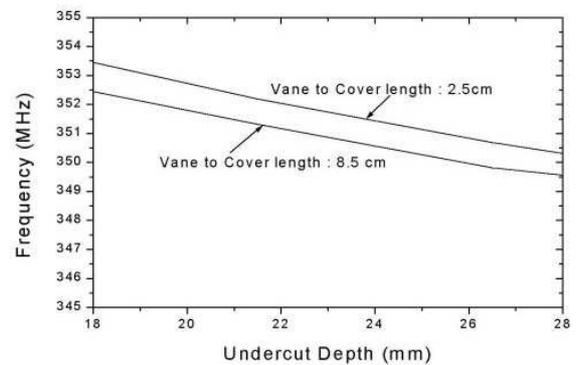

Figure 6. Plot of the resonant frequency versus the depth of the undercut.

## 4 PRESENT STATUS

The design of the 3MeV RFQ has been completed. The cold model with aluminium has been fabricated and tested. The 3MeV RFQ with OFHC is being fabricated. As a test bed for 3MeV RFQ, the design, construction, electrical test, and vacuum test of the 0.45MeV RFQ have been finished. The rf tuner has been fabricated and was tested.

## 5 ACKNOWLEDGMENT

We are grateful to Dr. Kazuo Hasegawa at JAERI, B.H. Lee and L.H. Whang at Vitzrotech Co. Ltd., and D.S. Lim at DUE Co. This work has been supported by the Korean Ministry of Science and Technology (MOST).